\documentclass[article,12pt,showkeys,showpacs] {revtex4}
\usepackage{amsfonts}
\usepackage{graphicx, amsmath}
\begin{document}

\title[Short Title]{One-step implementation of multiqubit phase gate with one control qubit and multiple target qubits in coupled cavities}
\author{Hong-Fu Wang\footnote{E-mail: hfwang@ybu.edu.cn}}
\affiliation{Department of Physics, College of Science, Yanbian
University, Yanji, Jilin 133002, People's Republic of China}
\author{Ai-Dong Zhu}
\affiliation{Department of Physics, College of Science, Yanbian
University, Yanji, Jilin 133002, People's Republic of China}
\author{Shou Zhang}
\affiliation{Department of Physics, College of Science, Yanbian
University, Yanji, Jilin 133002, People's Republic of China}
\begin{abstract}
We propose a one-step scheme to implement a multiqubit controlled phase gate of one qubit simultaneously controlling multiple qubits with three-level atoms at distant nodes in coupled cavity arrays. The selective qubit-qubit couplings are achieved by adiabatically eliminating the atomic excited states and photonic states and the required phase shifts between the control qubit and any target qubit can be realized through suitable choices of the parameters of the external fields. Moreover, the effective model is robust against decoherence because neither the atoms nor the field modes during the gate operation are excited, leading to a useful step toward scalable quantum computing networks.
\end{abstract}

\maketitle \section{Introduction and Motivation}\label{sec0}
It has been shown that any multiqubit gate can be decomposed into two classes of elementary quantum gates, namely, universal two-qubit controlled phase gate and one-qubit unitary gate, which are the basic building blocks of a quantum computer. However, the procedure of decomposing multiqubit gates into the elementary gates usually becomes very complicated as the number of qubits increases when using the conventional gate decomposition method. To reduce the complexity of the physical realization of practical quantum computing and quantum information processing, the direct implementation of multiqubit logic gates with multiple control qubits~\cite{HKPRA0470, LBHPRA0572, AGPRA0571, XZMYGPRA0673, CMKPRA0673, XYGPRA0674, YXGPRA077503, SHISPRA0775, SPRA1387} or multiple target qubits~\cite{CYFPRA1081, CSFPRA1082} is thus very important.

In recent years, considerable theoretical effort has been devoted to a class of coupled cavity models, which typically describe a series of optical cavities, each containing one or more atoms, photons are permitted to hop between the cavities. The coupled cavity arrays are promise to overcome the problem of individual addressability and have several interesting potential applications, including quantum information processing and simulations of quantum strongly correlated many-body systems. Theoretical studies on quantum quantum computing and quantum information processing have been put forwarded for the use of the atom-light interaction in coupled microcavity arrays~\cite{MFMPRL0799, DMSPRA0776, DANJP0810, CEMPRA0878, ECMPRA0877, JDSPRA0878, PYQGPRA0979, JYHAPL1096, ZZXXGPRA1081, LXZHSPRA1184}. In this paper, we propose a scheme for implementing a multiqubit phase gate with one control qubit and multiple target qubits with three-level atoms at distant nodes in coupled cavity arrays. This type of multiqubit controlled phase gate is essentially equivalent to $n$ two-qubit controlled phase gates, each having a shared control qubit (qubit 1) but a different target qubit (qubits $2,3,\ldots, N$). In the scheme, the selective off-resonant qubit-qubit coupling between the ground states of two atoms, induced by the external fields and the cavity modes, leads to an effective phase shift between the control qubit and any target qubit by choosing the parameters of the external fields appropriately. The scheme has the following merits: (i) it can be accomplished only in one step, which greatly simplifies the experimental realization and reduces the total gate time; (ii) it does not require the individual addressing of the trapped atoms; (iii) multiqubit gate operation among spatially separated quantum nodes is very meaningful for distributed quantum computing and quantum communication networks. Moreover, it would be an important step toward efficiently constructing quantum circuits and quantum algorithms.

\begin{figure}
\includegraphics[width=6.3in]{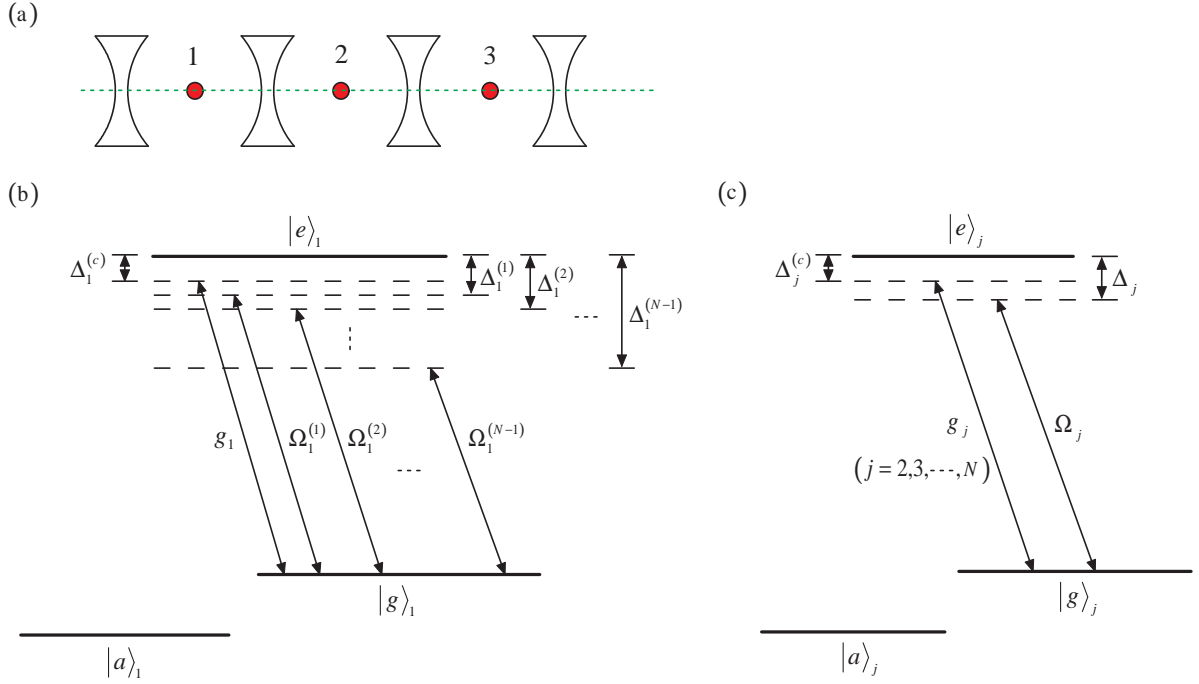}\caption{(a) A one-dimensional (1D) array of coupled cavities with one three-level atom in each cavity. (b) The level configuration and excitation scheme of atom 1. The transition $|g\rangle_1\leftrightarrow|e\rangle_1$ is coupled to the cavity mode with the coupling strength $g_1$ and is driven by $N-1$ classical fields with the Rabi frequencies $\Omega_1^{(k)}$ $(k=1,2,\ldots,N-1)$. (c) The level configuration and excitation scheme of atom $j$ $(j=2,3,\ldots,N)$. The transition $|g\rangle_j\leftrightarrow|e\rangle_j$ is coupled to the cavity mode with the coupling strength $g_j$ and is driven by a classical field with the Rabi frequency $\Omega_j$.}
\end{figure}

\section{Model and the Effective Hamiltonian}\label{sec1}
We consider an array of cavities that are coupled via exchange of photons with one three-level atom in each cavity, as sketched in Fig.~2(a). Such a model can be constructed in several kinds of physical systems such as superconducting stripline resonators~\cite{ADALRJSSRN04431}, photonic crystal defects~\cite{BSTYNM054}, and microtoroidal cavity arrays~\cite{STKKEHPRA0571}. Each atom has one excited state $|e\rangle$, and two ground states, $|g\rangle$ and $|a\rangle$. The transition $|g\rangle_i\rightarrow|e\rangle_i$ ($i=1,2,\ldots,N$) is coupled to the cavity mode with the coupling strength $g_i$ and detuning $\Delta_i^{(c)}$. On the other hand, the transition $|g\rangle_1\rightarrow|e\rangle_1$ for atom 1 is driven by $N-1$ classical laser fields with the Rabi frequencies $\Omega_1^{(k)}$ and detunings $\Delta_1^{(k)}$ ($k=1,2,\ldots,N-1$) (Fig.~2(b)). The transition $|g\rangle_j\rightarrow|e\rangle_j$ for atom $j$ ($j=2,3,\ldots,N$) is driven by a classical laser field with the Rabi frequency $\Omega_j$ and detuning $\Delta_j$ (Fig.~2(c)).
In the interaction picture, the Hamiltonian describing the atom-field interaction is
\begin{eqnarray}\label{e1}
\hat{H}_{\rm int}=\hat{H}_1+\hat{H}_2,
\end{eqnarray}
where
\begin{eqnarray}\label{e2}
\hat{H}_1&=&J_c\sum\limits_{j=1}^N\left(\hat{a}_j^\dag \hat{a}_{j+1}+\hat{a}_j \hat{a}_{j+1}^\dag\right),\cr\cr
\hat{H}_2&=&\left[\sum\limits_{j=1}^Ng_j \hat{a}_je^{i\Delta_j^{(c)}t}|e\rangle_j{_j}\langle g|+\sum\limits_{m=1}^{N-1}\Omega_1^{(m)}e^{i\Delta_1^{(m)}t}|e\rangle_1{_1}\langle g|+\sum\limits_{n=2}^N\Omega_ne^{i\Delta_nt}|e\rangle_n{_n}\langle g|\right]+{\rm H.c.},
\end{eqnarray}
and $J_c$ is the cavity-cavity hopping strength. Consider the periodic boundary conditions $\hat{a}_{N+1}=\hat{a}_1$ and take advantage of Fourier transformation to diagonalize the photon coupling terms, introducing the nonlocal bosonic modes and defining
\begin{eqnarray}\label{e3}
\hat{a}_j=\frac{1}{\sqrt{N}}\sum\limits_{k=1}^Ne^{-i\frac{2\pi}{N}jk}\hat{b}_k,
\end{eqnarray}
then the Hamiltonian $\hat{H}_1$ and $\hat{H}_2$ can be rewritten as
\begin{eqnarray}\label{e4}
\hat{H}_1&=&\sum\limits_{k=1}^N\omega_k \hat{b}_k^\dag \hat{b}_k,\cr\cr
\hat{H}_2&=&\left[\frac{1}{\sqrt{N}}\sum\limits_{j=1}^N\sum\limits_{k=1}^Ng_j \hat{b}_ke^{-i\frac{2\pi}{N}jk}e^{i\Delta_j^{(c)}t}|e\rangle_j{_j}\langle g|+\sum\limits_{m=1}^{N-1}\Omega_1^{(m)}e^{i\Delta_1^{(m)}t}|e\rangle_1{_1}\langle g|
\right.\cr\cr&&\left.+\sum\limits_{n=2}^N\Omega_ne^{i\Delta_nt}|e\rangle_n{_n}\langle g|\right]+{\rm H.c.},
\end{eqnarray}
where $\omega_k=2J_c\cos(\frac{2\pi}{N}k)$. We now go into a new frame by defining $H_1$ as a free Hamiltonian and perform the transformation $e^{iH_1t}$, obtaining
\begin{eqnarray}\label{e5}
\hat{H}^\prime&=&\left[\frac{1}{\sqrt{N}}\sum\limits_{j=1}^N\sum\limits_{k=1}^Ng_j \hat{b}_ke^{-i\frac{2\pi}{N}jk}e^{i\left(\Delta_j^{(c)}-\omega_k\right)t}|e\rangle_j{_j}\langle g|+\sum\limits_{m=1}^{N-1}\Omega_1^{(m)}e^{i\Delta_1^{(m)}t}|e\rangle_1{_1}\langle g|
\right.\cr\cr&&\left.+\sum\limits_{n=2}^N\Omega_ne^{i\Delta_nt}|e\rangle_n{_n}\langle g|\right]+{\rm H.c.}.
\end{eqnarray}
Under the conditions $\left|\Delta_j^{(c)}-\omega_k\right|\gg\frac{1}{\sqrt{N}}g_j$, $\Delta_1^{(m)}\gg\Omega_1^{(m)}$, and $\Delta_n\gg\Omega_n$, the upper level $|e\rangle_j$ can be adiabatically eliminated, leading to
\begin{eqnarray}\label{e6}
\hat{H}^{\prime\prime}&=&-\sum\limits_{k=1}^{N}\left[\sum\limits_{m=1}^{N-1}\xi_{m,k}\hat{b}_ke^{-i\frac{2\pi}{N}k}e^{i\left(\Delta_1^{(c)}
-\omega_k-\Delta_1^{(m)}\right)t}|g\rangle_1{_1}\langle g|\right.\cr\cr&&\left.+\sum\limits_{n=2}^{N}\zeta_{n,k}\hat{b}_ke^{-i\frac{2\pi}{N}nk}e^{i\left(\Delta_n^{(c)}
-\omega_k-\Delta_n\right)t}|g\rangle_n{_n}\langle g|+{\rm H.c.}\right]\cr\cr&&-\left(\sum\limits_{m=1}^{N-1}\eta_m|g\rangle_1{_1}\langle g|+\sum\limits_{n=2}^{N}\mu_n|g\rangle_n{_n}\langle g|+\sum\limits_{l=1}^{N}\sum\limits_{k=1}^{N}\chi_{l,k}\hat{b}_k^\dag \hat{b}_k|g\rangle_l{_l}\langle g|
\right),
\end{eqnarray}
where
\begin{eqnarray}\label{e7}
\eta_m&=&\frac{(\Omega_1^{(m)})^2}{\Delta_1^{(m)}},~~~\mu_n~=~\frac{\Omega_n^2}{\Delta_n},\cr\cr\chi_{l,k}&=&\frac{g_l^2}{N\left(\Delta_l^{(c)}-\omega_k\right)},
\cr\cr\xi_{m,k}&=&\frac{g_1\Omega_1^{(m)}}{2\sqrt{N}}\left(\frac{1}{\Delta_1^{(c)}-\omega_k}+\frac{1}{\Delta_1^{(m)}}\right),
\cr\cr\zeta_{n,k}&=&\frac{g_n\Omega_n}{2\sqrt{N}}\left(\frac{1}{\Delta_n^{(c)}-\omega_k}+\frac{1}{\Delta_n}\right).
\end{eqnarray}
The first two terms in Eq.~{(\ref{e6})} are the coupling between the bosonic
modes $\hat{b}_k$ and the classical field assisted by the atoms. The last three terms are the Stark shifts for the level $|g\rangle_j$ that are induced by the bosonic modes $\hat{b}_k$ and the classical pulse, respectively.
In the case of $\left|\Delta_1^{(c)}-\omega_k-\Delta_1^{(m)}\right|\gg\xi_{m,k}$ and $\left|\Delta_n^{(c)}
-\omega_k-\Delta_n\right|\gg\zeta_{n,k}$, the bosonic modes do not exchange quanta with the atomic system, the bosonic modes are only virtually excited and any two atoms interfere with each other during the whole interaction process. The effective Hamiltonian is then given by
\begin{eqnarray}\label{e8}
\hat{H}_{\rm eff}&=&\sum\limits_{k=1}^{N}\left\{\left[\sum\limits_{p=2}^N\sum\limits_{q=2,q\neq p}^{N}\Gamma_{p,q,k}e^{i\left(\Delta_p^{(c)}-\Delta_q^{(c)}
-\Delta_p+\Delta_q\right)t}|g\rangle_p{_p}\langle g|\otimes|g\rangle_q{_q}\langle g|\right.\right.\cr\cr
&&\left.+\sum\limits_{m=1}^{N-1}\sum\limits_{n=2}^{N}\left.\Lambda_{m,n,k}e^{i\left(\Delta_n^{(c)}-\Delta_1^{(c)}-\Delta_n
+\Delta_1^{(m)}\right)t}|g\rangle_1{_1}\langle g|\otimes|g\rangle_n{_n}\langle g|+{\rm H.c.}\right]\right\}\cr\cr&&+\sum\limits_{m=1}^{N-1}\left(\sum\limits_{k=1}^{N}\theta_{m,k}-\eta_m\right)|g\rangle_1{_1}\langle g|+\sum\limits_{n=2}^{N}\left(\sum\limits_{k=1}^{N}\vartheta_{n,k}-\mu_n\right)|g\rangle_n{_n}\langle g|\cr\cr&&-\sum\limits_{l=1}^{N}\sum\limits_{k=1}^{N}\chi_{l,k}\hat{b}_k^\dag \hat{b}_k|g\rangle_l{_l}\langle g|,
\end{eqnarray}
with
\begin{eqnarray}\label{e9}
\theta_{m,k}&=&\frac{\xi_{m,k}^2}{\Delta_1^{(c)}-\omega_k-\Delta_1^{(m)}},~~~
\vartheta_{n,k}~=~\frac{\zeta_{n,k}^2}{\Delta_n^{(c)}-\omega_k-\Delta_n},\cr\cr
\Gamma_{p,q,k}&=&\frac{\zeta_{p,k}\zeta_{q,k}e^{-i\frac{2\pi}{N}(p-q)k}}{2}\left(\frac{1}{\Delta_p^{(c)}-\omega_k-\Delta_p}
+\frac{1}{\Delta_q^{(c)}-\omega_k-\Delta_q}\right),\cr\cr
\Lambda_{m,n,k}&=&\frac{\xi_{m,k}\zeta_{n,k}e^{-i\frac{2\pi}{N}(n-1)k}}{2}\left(\frac{1}{\Delta_1^{(c)}-\omega_k-\Delta_1^{(m)}}
+\frac{1}{\Delta_n^{(c)}-\omega_k-\Delta_n}\right),
\end{eqnarray}
where $\zeta_{\nu,k} (\nu=p,q,n)$ in Eq.~(9) is denoted by Eq.~(7). As the quantum number of the bosonic modes is conserved during the interaction, they will remain in the vacuum state if they are initially in the vacuum state. Choosing the detunings suitably so that
\begin{eqnarray}\label{e10}
&&\Delta_n^{(c)}-\Delta_1^{(c)}-\Delta_n
+\Delta_1^{(m)}=0, ~~(m=n-1, ~n\in\{2,3,\ldots,N\}),\cr\cr&&
\left|\Delta_n^{(c)}-\Delta_1^{(c)}-\Delta_n
+\Delta_1^{(m)}\right|\gg \left|\sum\limits_{k=1}^{N}\Lambda_{m,n,k}\right|, ~~(m\neq n-1),\cr\cr&&
\left|\Delta_p^{(c)}-\Delta_q^{(c)}-\Delta_p+\Delta_q\right|\gg \left|\sum\limits_{k=1}^{N}\Gamma_{p,q,k}\right|, ~~(p,q\in\{2,3,\ldots,N\},~p\neq q),
\end{eqnarray}
then the effective Hamiltonian reduces to
\begin{eqnarray}\label{e11}
\hat{H}_{\rm eff}^\prime&=&\sum\limits_{j=2}^{N}\left(\zeta^\prime_{1,j}|g\rangle_1{_1}\langle g|+\xi^\prime_{j}|g\rangle_j{_j}\langle g|+\Lambda_{1,j}^\prime|g\rangle_1{_1}\langle g|\otimes|g\rangle_j{_j}\langle g|\right),
\end{eqnarray}
with
\begin{eqnarray}\label{e12}
\zeta^\prime_{1,j}&=&\sum\limits_{k=1}^{N}\theta_{j-1,k}-\eta_{j-1},~~~\xi^\prime_{j}~=~\sum\limits_{k=1}^{N}\vartheta_{j,k}-\mu_j, \cr\cr\Lambda_{1,j}^\prime&=&\sum\limits_{k=1}^{N}\xi_{j-1,k}\zeta_{j,k}\cos\left[\frac{2\pi}{N}(j-1)k\right]
\left(\frac{1}{\Delta_1^{(c)}-\omega_k-\Delta_1^{(j-1)}}
+\frac{1}{\Delta_j^{(c)}-\omega_k-\Delta_j}\right),
\end{eqnarray}
where $\eta_{j-1}$, $\mu_j$, $\xi_{j-1,k}$, and $\zeta_{j,k}$ are denoted by Eq.~({\ref{e7}}) and $\theta_{j-1,k}$ and $\vartheta_{j,k}$ are denoted by Eq.~({\ref{e9}}), respectively. The last term in Eq.~({\ref{e11}}) describes the coupling between atoms 1 and $j$ ($j=2,3,\ldots,N$) mediated by the bosonic modes
and the classical pulses.

\section{Implementation of Multiqubit Phase Gate With One Control Qubit and Multiple Target Qubits}\label{sec2}
To implement quantum computation, the two long-lived levels $|a\rangle$ and $|g\rangle$ represent the states of the qubits, which correspond the logical zero and one states, respectively, $|0\rangle\equiv |a\rangle$ and $|1\rangle\equiv |g\rangle$. Due to the the virtual excitation of the atoms in the interaction, the atom 1 and any atom $j$ will undergo an energy shift, leading to the evolution
\begin{eqnarray}\label{e13}
|a\rangle_1|a\rangle_j&\rightarrow&|a\rangle_1|a\rangle_j,~~~~~~~~~~~
|a\rangle_1|g\rangle_j~\rightarrow~e^{-i\varphi_j}|a\rangle_1|g\rangle_j,\cr\cr
|g\rangle_1|a\rangle_j&\rightarrow&e^{-i\psi_{1,j}}|g\rangle_1|a\rangle_j,~~~
|g\rangle_1|g\rangle_j~\rightarrow~e^{-i(\varphi_j+\psi_{1,j}+\phi_{1,j})}|g\rangle_1|g\rangle_j,
\end{eqnarray}
where $\varphi_j=\xi^\prime_jt$, $\psi_{1,j}=\zeta^\prime_{1,j}t$, and $\phi_{1,j}=\Lambda_{1,j}^\prime t$. After the performance of the one-qubit phase shifts $|g\rangle_1\rightarrow e^{i\psi_{1,j}}|g\rangle_1$
and $|g\rangle_j\rightarrow e^{i\varphi_j}|g\rangle_j$, a conditional phase shift $\phi_{1,j}$, which is controllable via the effective interaction time $t$ and the corresponding parameters $\Lambda_{1,j}^\prime$, is produced if and only if atoms 1 and $j$ are in the state $|g\rangle$. While for the state $|g\rangle_m|g\rangle_n$ of other any pair of atoms $m$ and $n$ ($m,n=2,3,\ldots,N$), no conditional phase shift is generated. Therefore, for the computational basis states $\{|s^1\rangle_1|s^2\rangle_2|s^3\rangle_3\otimes\cdots\otimes|s^{n-1}\rangle_{N-1}|s^n\rangle_N\}$ ($s^i=a,g$; $i=1,2,\ldots,n$) consisting of $N$ atoms, after the qubit-qubit couplings and a series of one-qubit phase shift operations on the state $|g\rangle_k$ ($k=1,2,\ldots,N$) and with the choice of $\phi_{1,j}=\Lambda_{1,j}^\prime t=\pi$, one can obtain
\begin{eqnarray}\label{e14}
|a\rangle_1|s^2\rangle_2|s^3\rangle_3\otimes\cdots\otimes|s^{n-1}\rangle_{N-1}|s^n\rangle_N~~~~~~~~~~~~~~~~~~~~~~~~~~
~~~~~~~~~~~~~~~~~~~~~~~~~~~~~~~\cr\cr
\longrightarrow|a\rangle_1|s^2\rangle_2|s^3\rangle_3
\otimes\cdots\otimes|s^{n-1}\rangle_{N-1}|s^n\rangle_N,~~~~~~~~~~~~~~~~~~\cr\cr
|g\rangle_1|s^2\rangle_2|s^3\rangle_3\otimes\cdots\otimes|s^{n-1}\rangle_{N-1}|s^n\rangle_N~~~~~~~~~~~~~~~~~~~~~~~~~~
~~~~~~~~~~~~~~~~~~~~~~~~~~~~~~~\cr\cr\longrightarrow
e^{-i\sum\limits_{j=2}^{N}(1-g\oplus s^j)\pi}|g\rangle_1|s^2\rangle_2|s^3\rangle_3\otimes\cdots\otimes|s^{n-1}\rangle_{N-1}|s^n\rangle_N,
\end{eqnarray}
where $g\oplus g=0$ and $g\oplus a=1$. Therefore, a multiqubit controlled phase gate of one qubit simultaneously controlling multiple qubits is achieved if and only if the atom 1 is in the state $|g\rangle_1$.

\section{Analysis and discussion}\label{sec3}
We now give a brief analysis and discussion for some practical issues in relation to the experimental feasibility of the proposed scheme. For simplicity, we consider the case with $N=3$ and set $J_c=0.5g$, $g_1=g_2=g_3=g$, $\Delta_1^{(c)}=\Delta_2^{(c)}=\Delta_3^{(c)}=20g$, $\Omega_1^{(1)}=\Omega_1^{(2)}=\Omega_2=\Omega_3=g$, $\Delta_1^{(1)}=\Delta_2=18g$, and $\Delta_1^{(2)}=\Delta_3=21.2842g$. Then we have
\begin{eqnarray}\label{e15}
\Lambda^\prime_{1,2}=\Lambda^\prime_{1,3}=\Lambda_3^\prime&=&\sum\limits_{k=1}^{3}\frac{\cos(\frac{2k\pi}{3})}{6\left[2-\cos(\frac{2k\pi}{3})\right]}
\left(\frac{1}{20-\cos(\frac{2k\pi}{3})}+\frac{1}{18}\right)^2\cr\cr
&=&1.225\times 10^{-3}g,
\end{eqnarray}
and the time needed to complete the three-qubit controlled phase gate operation is $t_3=\pi/\Lambda_3^\prime=2.56457\times 10^3/g$, and all the restrictive conditions in the cases of large detunings are well satisfied. The probability that the atoms undergo a transition to the excited state due to the qubit-qubit coupling with the classical fields is
\begin{eqnarray}\label{e16}
p_e&=&\frac{1}{8}\times\left[\frac{3}{4}\times\frac{(\Omega_1^{(1)})^2+(\Omega_1^{(2)})^2}{(\Delta_1^{(c)})^2}
+\frac{13}{8}\times\frac{(\Omega_2)^2}{(\Delta_2^{(c)})^2}+\frac{3}{8}\times\frac{(\Omega_3)^2}{(\Delta_3^{(c)})^2}\right]\cr\cr
&=&1.09375\times 10^{-3}.
\end{eqnarray}
Meanwhile, the probability that the field mode is excited due to the qubit-qubit coupling is
\begin{eqnarray}\label{e17}
p_c&=&\frac{3}{4}\times\left[\sum\limits_{k=1}^{3}\left(\frac{\xi_{1,k}}{\Delta_1^{(c)}-\omega_k-\Delta_1^{(1)}}\right)^2
+\sum\limits_{k=1}^{3}\left(\frac{\xi_{2,k}}{\Delta_1^{(c)}-\omega_k-\Delta_1^{(2)}}\right)^2\right]\cr\cr
&&+\frac{13}{8}\times\left[\sum\limits_{k=1}^{3}\left(\frac{\zeta_{2,k}}{\Delta_2^{(c)}-\omega_k-\Delta_2}\right)^2\right]
+\frac{3}{8}\times\left[\sum\limits_{k=1}^{3}\left(\frac{\zeta_{3,k}}{\Delta_3^{(c)}-\omega_k-\Delta_3}\right)^2\right]\cr\cr
&=&5.98033\times 10^{-3}.
\end{eqnarray}
Therefore, the effective Hamiltonian $\hat{H}_{\rm eff}^\prime$ in Eq.~({\ref{e11}}) is valid. Furthermore, the effective decoherence rates due to the atomic spontaneous emission and the field decay are $\gamma_e=p_e\gamma$ and $\kappa_c=p_c\kappa$, with $\gamma$ and $\kappa$ being the decay rates for the atomic excited state and the field modes, respectively. If we take the parameters $\gamma\sim \kappa \sim 3\times 10^{-3}g$, which were predicted to be available~\cite{STKKEHPRA0571, SCFPRA1082}, the corresponding gate fidelity is about $F\simeq 1-(\gamma_e+\kappa_c)t\simeq 95\%$.
When we set $N=4$, with the choices of $J_c=0.5g$, $g_1=g_2=g_3=g_4=g$, $\Omega_1^{(1)}=\Omega_1^{(2)}=\Omega_1^{(3)}=\Omega_2=\Omega_3=\Omega_4=g$, $\Delta_1^{(c)}=\Delta_2^{(c)}=\Delta_3^{(c)}=\Delta_4^{(c)}=20g$, $\Delta_1^{(1)}=\Delta_2=18g$, $\Delta_1^{(2)}=\Delta_3=18.34g$, and $\Delta_1^{(3)}=\Delta_4=21.7492g$, we have
\begin{eqnarray}\label{e15}
\Lambda^\prime_{1,2}=\Lambda^\prime_{1,3}=\Lambda^\prime_{1,4}=\Lambda_4^\prime=1.0195\times 10^{-3}g,
\end{eqnarray}
and the time needed to complete the four-qubit controlled phase gate operation is $t_4=\pi/\Lambda_4^\prime=3.0815\times 10^3/g$. As reported in cavity QED experiment in Ref.~\cite{ARKAJHPRL0493}, the coupling strength can be achieved as $g=2\pi\times 34~{\rm MHz}$. Therefore, the time required to implement three-qubit controlled phase gate is on the order of $t_3=1.20048\times 10^{-5}$~s, and $t_4=1.44246\times 10^{-5}$~s for four-qubit phase gate. In the experiments, the decay time of the cavity is $t_c=3.0\times 10^{-2}$~s~\cite{AGSPMJSS00288, PAPSTMJSPRL0289}, which is longer than the required times. Even when $N=200$, by setting $J_c=0.5g$, $g_l=g$, $\Delta_l^{(c)}=20g$ ($l=1,2,\ldots,N$), $\Omega_1^{(m)}=\Omega_n=g$ ($m=1,2,\ldots, N-1$; $n=2,3,\ldots, N$), $\Delta_1^{(1)}=\Delta_2=18g$, we obtain $\Lambda_{200}^\prime=9.45658\times 10^{-4} g$ and the time required to achieve the phase gate operation is $t_{200}=\pi/\Lambda_{200}^\prime=1.5551\times 10^{-5}$~s. Therefore, based on the current cavity QED techniques, the proposed scheme might be experimentally realizable.

\section{Conclusions}\label{sec4}
In conclusion, we have proposed an efficient scheme to realize the selective coherent coupling between the control qubit and any target qubit in a 1D quantum network. We consider the scheme in a 1D coupled cavity system with three-level atoms at distant nodes trapped in separated cavities. It is unnecessary to
utlize a single-photon source or inject a single-photon into an optical cavity. With a suitable choice of the system parameters, a multiqubit phase gate with one control qubit and multiple target qubits can be directly achieved, which greatly simplifies the experimental realization and reduces the total gate time. Furthermore, the interaction among the atoms we use to implement the multiqubit phase gate can also be realized in other systems with three-level configuration.

\begin{center}
{\small {\bf ACKNOWLEDGMENTS}}
\end{center}

This work is supported by the National Natural Science Foundation
of China under Grant Nos. 11264042, 61068001, and 11165015; the China Postdoctoral
Science Foundation under Grant No. 2012M520612; the Program for Chun Miao Excellent Talents of Jilin Provincial Department of Education under Grant No. 201316; and the Talent Program of Yanbian University of China under Grant No. 950010001.

\end{document}